# High performance computing network for cloud environment using simulators

Ajith Singh. N[1] and M. Hemalatha[2]

[1]Ph.D, Research Scholar (CS), Karpagam University, Coimbatore, India
[2] Prof & Head, Department of Software Systems & Research, Karpagam University, Coimbatore, India

Email: ajithex@gmail.com[1], hema.bioinf@gmail.com[2]

## Abstract

**Cloud computing is the next generation computing. Adopting the cloud computing is like signing up new form of a website. The GUI which controls the cloud computing make is directly control the hardware resource and your application. The difficulty part in cloud computing is to deploy in real environment. Its' difficult to know the exact cost and it's requirement until and unless we buy the service not only that whether it will support the existing application which is available on traditional data center or had to design a new application for the cloud computing environment. The security issue, latency, fault tolerance are some parameter which we need to keen care before deploying, all this we only know after deploying but by using simulation we can do the experiment before deploying it to real environment. By simulation we can understand the real environment of cloud computing and then after it successful result we can start deploying your application in cloud computing environment. By using the simulator it will save us lots of time and money.**

## Introduction

Cloud Computing [1] [4] a marketing term which is also known as utility computing deliver the service as software, platform and infrastructure as a service in pay-as-you-go model to consumers. Berkeley report says on this services as "Cloud computing, the long held dream of computing as a utility, has the potential to transform a large part of the IT industry, making software even more attractive as a service.

Moving an existing in-house application or new application to the cloud will almost surely have some trade-offs in terms of performance [7][8][9]. Existing applications were not designed for cloud environment. Organizations considering moving to cloud computing will certainly have to think in costs for improving the network infrastructure required to run applications in the cloud [10]. On the other side, bandwidth continues to increase and approaches such as dynamic caching, compression, pre-fetching and other related web-acceleration technologies can effect in major performance improvements for end users, often exceeding 50%. The application has to design or redesign in such a way that it will support cloud computing in order to achieve maximum performance [11]. Azure is a Microsoft cloud platform to build, host and scale web application through Microsoft data centers [12].

Cloud computing growth has taken all the attention of various communities like researches, student, business, consumer and government organization. Big data is the main reason for coming of cloud computing in the show, everyday lots of data in the size of PETA bytes are uploaded in the digital world which required lots of storage and computing resources.

Cloud based and traditional web application which in short includes social networking [14], web hosting, content delivery, and real time instrumented data processing. Each of these application types different composition, configuration, and deployment requirements. Measuring the performance of scheduling and allocation policies in a real Cloud environment for different application and service models under different conditions is extremely challenging issue because (i) clouds exhibit varying demand, supply patterns, system size and (ii) users have heterogeneous and competing QoS requirements. Deploying the cloud computing in real infrastructure such as Amazon EC2 [15], Google [16], limits the experiments to the scale of the infrastructure, and make the reproduction of results an extremely difficult undertaking. The main reason for this being the conditions prevailing in the Internet-based environments are beyond the control of developers of resource allocation and application scheduling algorithms.

Alternative to real infrastructure is simulation tools that open to study before deploying in real environment. Simulation is cost free just we need to download simulation based on our work or design it with any programming language. It works as real environment which help us to study the real architecture of the cloud computing. While simulation we can study the exact real environment before putting our application online, by simulation we can study the load, fault tolerance security latency and budget. By simulation we can test the service, experiment with different workload and resource performance. Cloud computing is a large scale distributed systems, simulation and many tool has been developed to study the distributed systems. Some of these are GridSim, MicroGrid, GangSim, SimGrid and CloudSim. The remaining of the paper goes on discussing the available simulator which is available for cloud computing and a experiment on CloudAnalyst using Round robin load balancing algorithm.

## Overview of Simulator

### A. GridSim

Peer-to-peer system or Grid Computing system are of millions of node which is used for large scale distributed systems where workload is distributed among these node. GridSim [13] toolkit was developed by Buyya et al.

Doing an experiment in such a large environment is difficult. GridSim [13] simulator is designed and developed to cope with these properties then to reach intense scalability and to hold dynamism by avoiding the overhead of co-ordination of real resources.

### Overview of GridSim functionalities: [13]

- Incorporates failures of Grid resources during runtime.
- New allocation policy can be made and integrated into the GridSim Toolkit, by extending from AllocPolicy class.
- Has the infrastructure or framework to support advance reservation of a grid system.
- Incorporates a functionality that reads workload traces taken from supercomputers for simulating a realistic grid environment.
- Incorporates an auction model into GridSim.
- Incorporates a data grid extension into GridSim.
- Incorporates a network extension into GridSim. Now, resources and other entities can be linked in a network topology.
- Incorporates a background network traffic functionality based on a probabilistic distribution. This is useful for simulating over a public network where the network is congested.
- Incorporates multiple regional GridInformationService (GIS) entities connected in a network topology. Hence, you can simulate an experiment with multiple Virtual Organizations (VOs).
- Adds ant build file to compile GridSim source files.

A view of GridSim Job Scheduling report

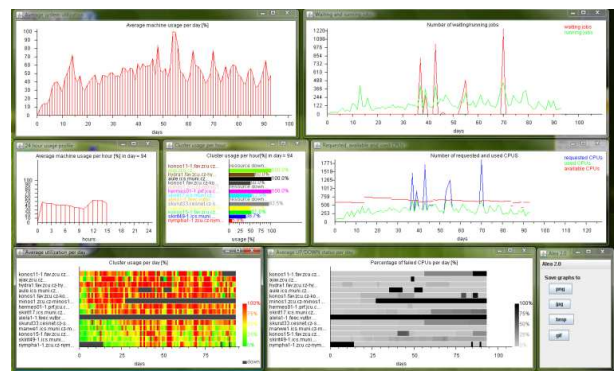

**Fig (1) Image Source:**
http://www.fi.muni.cz/~xklusac/alea/alea-gui.png

### B. CloudSim

Cloud computing [1][4] is the emerging technology which emphasizes commercial computing. Cloud is a platform providing dynamic resource pools, virtualization & high availability [5]. It's the concept implemented to overcome our regular computing problem like hardware software resource availability and related aspects. Cloud computing makes it easy to have high performance computing. Cloud computing is a service which is easily available on market when you want it you can start the service and when you don't want you stop the service and you pay it for what you use, you need is a thin client or a laptop to access the internet. To host any application in such an environment a test is compulsory. But to test a cloud based application is highly cost in real environment need to spend large currency. A suitable alternate to real environment is simulator as mention before. Simulation environment will allow evaluation of different kinds if resource learning scenario under varying loads and pricing distribution.

CloudSim [6] is a framework developed by the GRIDS Laboratory at University of Melbourne at 2009 to –date is a complete simulation for the cloud computing where researches and industry developers can focus on specific system design issues that they want to investigate, without getting concerned about the low level details related to Cloud-based infrastructures and services.[6]

Overview of CloudSim functionalities [6]:

- support for modeling and simulation of large scale Cloud computing data centers
- support for modeling and simulation of virtualized server hosts, with customizable policies for provisioning host resources to virtual machines
- support for modeling and simulation of energy-aware computational resources
- support for modeling and simulation of data center network topologies and message-passing applications
- support for modeling and simulation of federated clouds
- support for dynamic insertion of simulation elements, stop and resume of simulation
- support for user-defined policies for allocation of hosts to virtual machines and policies for allocation of host resources to virtual machines

## C. CloudAnalyst

With the above simulator and many other it is easy to do simulation on large scale environment. But it would be easier if the tool is with visualization capability. The tool which will separate the simulation experiment and programming exercise. By this we can concentrate of simulation parameter rather than the technicalities of programming. By using cloud analyst it's easy to do simulation and the get result in graphical view which is easy to understand and print the report. It's ease of use, ability to define the simulation with a high degree of configurability and flexibility, graphical output, repeatability ease of extension/ technologies used in cloud analyst are java, Java Swing, CloudSim and SimJava. SimJava is the underlying simulation framework of CloudSim and some feature of SimJava are used directly in CloudAnalyst [13].

**Experiment on CloudAnalyst simulation with below parameter.**

Two Data center with OS as Linux and VMM as Xen is created. And their physical Hardware details are below
Memory: 20480 MB
Storage: 128000 MB
No of Processors: 4
Processor Speed 10000
VM Policy: Time Shared

Simulation duration is set to 60 min with 6 User bases for different location. Table 1. below is data provided in every user base.

| Name | Region | Requests /user per hr | Data size /request (bytes) | Peak hrs start (gmt) | Peak hrs End (gmt) | Avg peak users | Avg off peak users |
|---|---|---|---|---|---|---|---|
| UB1 | 0 | 15 | 100 | 15 | 17 | 450000 | 600000 |
| UB2 | 1 | 15 | 100 | 17 | 22 | 500000 | 300000 |
| UB3 | 2 | 15 | 100 | 13 | 20 | 200000 | 60000 |
| UB4 | 3 | 15 | 100 | 14 | 18 | 250000 | 10000 |
| UB5 | 4 | 15 | 100 | 16 | 24 | 100000 | 200000 |
| UB6 | 5 | 15 | 100 | 18 | 22 | 300000 | 5000 |

**Table: 1 User bases for different location**

**The output of the simulation is below.**

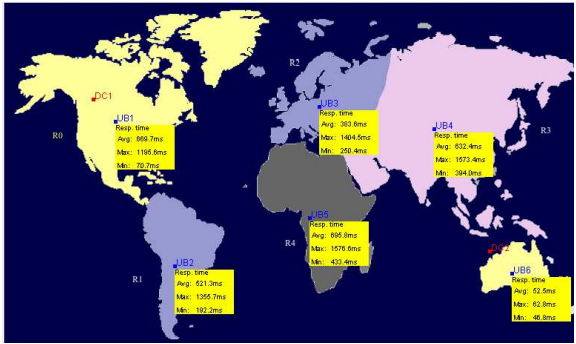

**Fig.2 Screen shot in geographical view**

Overall Response Time

|  | Avg (ms) | Min (ms) | Max (ms) |
|---|---|---|---|
| Overall Response Time: | 617.83 | 46.51 | 1576.30 |
| Data Center Processing Time: | 410.86 | 1.33 | 1108.95 |

**Table: 2 Overall Response Time**

Response time by the Region

| Userbase | Avg (ms) | Min (ms) | Max (ms) |
|---|---|---|---|
| UB1 | 669.46 | 70.743 | 1194.944 |
| UB2 | 521.726 | 192.155 | 1353.876 |
| UB3 | 383.684 | 250.361 | 1405.049 |
| UB4 | 628.765 | 394.043 | 1573.967 |
| UB5 | 695.896 | 433.372 | 1576.303 |
| UB6 | 52.37 | 46.511 | 62.779 |

**Table: 3 Response time by the region**

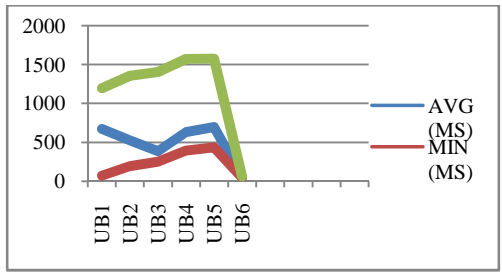

**Chart 1: Graphical view of response timbe by the region**

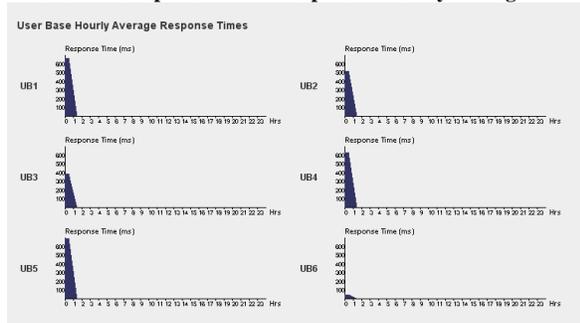

**Fig. 3 User Base Hourly Average Response TImes**

Data Center Request Servicing Times

| Data Center | Avg (ms) | Min (ms) | Max (ms) |
|---|---|---|---|
| DC1 | 412.605 | 2.502 | 1108.952 |
| DC2 | 2.954 | 1.327 | 4.085 |

**Table: 4 Data Center Request Servicing Times**

**Result:** With the above simulation we are able to define cost of such an environment in real world

**Cost**

Total virtual Machine Cost; $ 5.01
Total Data Transfer Cost: $172.16
Grand Total: $177.17

**Cost in Data Center wise.**

| Data Center | VM Cost | Data Transfer Cost | Total |
|---|---|---|---|
| DC1 | 2.503 | 0.735 | 3.238 |
| DC2 | 2.503 | 171.429 | 173.933 |

**Table: 5 Data Center Request Servicing Times**

## Conclusion & Future Work

We have discuss three simulator based on High performance computing network, GridSim for Grid Computing, CloudSim for Cloud Computing and CloudAnalyst for cloud environment cost wise. A small simulation with some parameter to understand the cost of using cloud computing in real world is experiment. From CloudAnalyst we are able to predict that a user wants to deploy on HPC network we can use a simulation to study it before deploying in real world. A simulation is done on CloudAnalyst is show us we much it going to cost if we want to deploy our application on cloud computing environment. Future work will be a comparative study on load balancing algorithm on CloudAnalyst and compare it result and find out which one is best load balancing algorithm in cloud computing environment.

**Acknowledgement**

We thank Karpagam University for motivating and encouraging doing our Research work in a Successful.